# FAIR: A Hadoop-based Hybrid Model for Faculty Information Retrieval System

Noopur Gupta
IIIT-Bhubaneswar,
Bhubaneswar, India
noopur2827@gmail.com

Rakesh K. Lenka
IIIT-Bhubaneswar,
Bhubaneswar, India
rakeshkumar@iiit-bh.ac.in

Rabindra K. Barik
KIIT University,
Bhubaneswar, India
rabindra.mnnit@gmail.com

Harishchandra Dubey
University of Texas
Dallas, USA
harishchandra.dubey@utdallas.edu

*Abstract*—In era of ever-expanding data and knowledge, we lack a centralized system that maps all the faculties to their research works. This problem has not been addressed in the past and it becomes challenging for students to connect with the right faculty of their domain. Since we have so many colleges and faculties this lies in the category of big data problem. In this paper, we present a model which works on the distributed computing environment to tackle big data. The proposed model uses apache spark as an execution engine and hive as database. The results are visualized with the help of Tableau that is connected to Apache Hive to achieve distributed computing.

## I. INTRODUCTION

In this era of internet, we can find everything online, which is true for finding faculties and colleges. But the major problem here is that this information is not centralized, i.e. you have to visit every college webpage and then to faculty page and then to their profile to see their research interest. Which is very time consuming and not at all efficient. We try to resolve this problem by developing a centralized system which will aid to find colleges and relevant faculties.

To deal with this problem first we need a centralized database which has all the information of all the faculties. Doing this will land us in the Big Data territory [1]. Imagine having information of all the colleges and its faculties at one place, how large the dataset would be and how time consuming will it be to run simple query. Dealing with this type of problem we require to build our model around tools which are capable in handling such mammoth amount of data, few of the tools are Apache Spark, Apache Hive [2].

## II. BIG DATA

Big Data is defined by its following 5 properties i.e. Volume, Variety, Veracity, Value, and Velocity [3]. The data growth has been exponential in the past few years, this statement can be supported by the statement made by Eric Schmidt (executive chairmen google) "From dawn of civilization until 2003, humankind generated five exabyte of data. Now we produce five exabyte of data every two days". Every company regardless of its industrial background stores its data for analysis to take smart decision for the company. The first two companies to face this big data challenge were Google and Facebook, and these were the companies who led to the foundation of big data tools that we use today. Big data is everywhere, it can be data collected by sensors all around the world or the live traffic on internet or in healthcare [4]. 204 million emails are send every minute, 2 million search queries are received by google in 2012 and this data doubled in 2014 etc, this large amount of data gathered from different data sources resulted in formation of big data [5]. Now to tackle this data related issue new tools or frameworks were required that can handle this large volume of data in an efficient and cost-effective way. Google first introduced MapReduce framework and with time news tools came in action, some of these tools are explained here [6].

## III. BIG DATA VISUALIZATION

Visualization has proved to be a great ally for finding useful patterns and information hidden in the data. It is not a new thing, since it has been used from decades to communicate information in a better and easier way [25]. A vast amount of data is generated every day, only 20% of this is structed rest is all unstructured data, visualizing data helps in exploring and analyzing data in easier way as it helps in understanding data, finding correlations, finding general trends in data etc. There exist some special visualization tools that are designed especially for handling big data like Tableau, Microsoft PowerBI, Gephie etc. in our model we used Tableau for visualization to do in-depth analysis [7].

## IV. ANATOMY OF BIG DATA TOOLS

Hadoop environment consists of many tools that helps in performing distributed and parallel computation. It helps in enabling scalability as one can store large volumes of data on commodity hardware's, it also provides fault tolerance, it can easily handle different data types and provides shared environment as it allows multiple jobs to execute simultaneously. In 2004 MapReduce framework was introduced by google, then in 2005 yahoo introduced Hadoop based on MapReduce framework. Today there are over 100 open source projects (tools) for big data and continuous to grow [8].

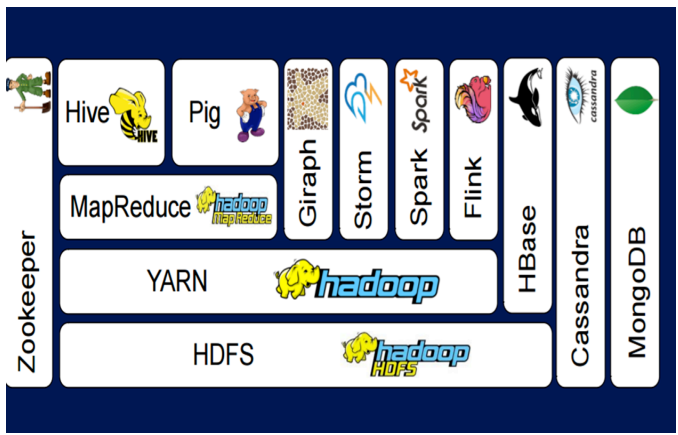

Figure 1: Hadoop Environment

*a) HDFS*

It stands for Hadoop distributed file system. HDFS is basically a storage system for big data, this file system resulted in foundation of several big data frameworks as it can provide scalability and a reliable storage. In today's era, there is no limit for data so a fixed size data warehouse can't serve the right purpose, data storage should be scalable and this scalability is provided by HDFS [9]. As size of our data increases, user can add more commodity hardware's to HDFS to increase capacity. A single large file will require a lot of space in storage so HDFS splits a large file into small chunks, then store these data chunks in multiple nodes, which helps in providing parallel access and computation. HDFS replicates file block (data chunks) on different nodes to prevent data loss, it maintains 3 copies of file blocks by default. To read or write a file in HDFS one should specify input file format and output file format respectively [10]. HDFS consists of two main components, they are name node and data node.

1) *Name Node:* It works on master node in the cluster, it stores metadata about the file like, it records the name of the file, location of file in directory, coordinates operations and mapping of data chunks in data node. There is also a secondary name node which acts as a backup for primary name node.
2) *Data Node:* It resides on slave node of the cluster, takes orders from the name node and act accordingly like file block deletion, replication etc. [11].

*b) YARN*

It works as a resource manager for Hadoop, yarn is added between HDFS and applications so that it can interact with applications and schedule resources for them. Because of Yarn other big data applications were built like giraph, storm, etc. yarn decides which node get what resources [12]. There are 3 main components in yarn they are:

1) *Node Manager:* it is present on each node in cluster and oversees that node.

2) *Application Master:* It communicates with the node manager for its task to be done and requests for resources from the resource manager.

3) *Container:* it holds all the resources that has been allocated to that node.

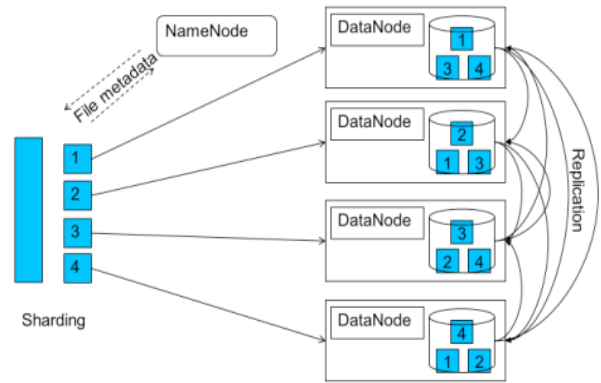

Figure 2: Structure of HDFS

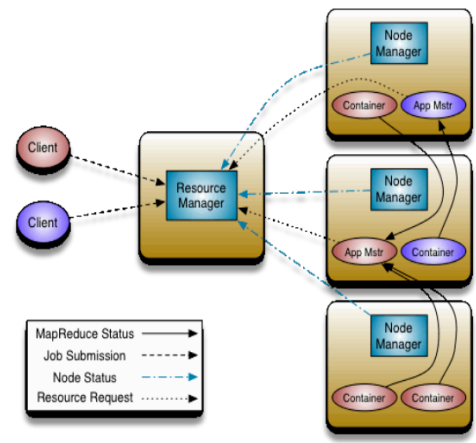

Figure 3: Structure of Yarn

*c) MapReduce*

It is a programming model that helps in parallel computation in uncomplicated way. It communicates with YARN for resource scheduling. To make this programming model work user need to provide two functions. A map () function is required which will be applied to each node in the cluster, this function reads the data and provides output in key – value pairs. Reduce () takes the key-value pairs provided by map (), reads it and tries to summarize elements in some manner [13].

*d) NoSQL Databases*
1) *Apache HBASE:* It is a non-relational distributed database and falls in category of column data model, it runs on the

top of HDFS and is scalable as user can easily add commodity clusters to store data. It helps in compressing the size of data on disk and can perform in-memory operations [14].
2) *Apache Cassandra:* it also falls under the category of column data model, can run with or without HDFS, it is an open source fork of a standalone database system. It is scalable and provides high availability [15].
3) *MangoDB*: It also belongs to NoSQL family and falls under Document-oriented database system category. It is free and open source platform, is scalable and provides high performance as well as high availability [16].

*e) Apache Pig*

It is a high-level platform used for data processing, as it is helpful in manipulating raw data. The language it uses is called pig Latin. Pig can take data from files, streams or any other sources by using user defined functions. It can easily perform ETL operations; it also has several inbuilt functions like average, min., max., absolute value, join, filter, sort etc. For more complex operations user can define its own function [17]. Pig can run locally as well as on other execution environments like it can run on the top of MapReduce or on the top of Tez.

*f) Apache Hive*

Hive is a data ware house that allows us to query and manage data. Data can be present in any distributed storage like HDFS, MapReduce, HBase, S3 etc. it can easily take data from all these means and creates a table for providing a structure to the data so that data can be managed easily and in which user can perform queries. The language it uses is called HiveQL, this query language is very like SQL thus allow user to use SQL like queries to access the data [18]. It can use different execution environments like MapReduce, spark or Tez.

*g) Apache Spark*

Spark is the hottest tool nowadays for distributed and parallel computation as it is much faster than any other tool in the environment and provides various libraries for better analysis of our data like for streaming, machine learning etc. Because of some demerits in MapReduce framework we required a new framework that can overcome the shortcomings of previous tools, thus Spark came in action. It provides a very rich programming interface and makes the workflow pipeline easier. It not only does faster batch processing but also does Real time processing, interactive data analysis which in turn makes faster decision making. The most important feature of spark is it performs in-memory computation, means your data will be loaded from disk into memory so that the data will not to be read from disk again and again as it will be cached in the memory, it will provide increase in the performance and that will be helpful in iterative algorithms like ML [19]. The languages through which one can work on spark are python, R, Scala, java. Java Virtual Machine is the Execution Engine used by Spark and it also acts as the interface for the rest of the Hadoop ecosystem. The two important things required for spark are distributed storage and cluster manager. For cluster management, it can run on yarn or apache mesas or standalone, for distributed storage it can use HDFS or Amazon S3. When spark is combined with Hive it is known as SHARK, and SHARK is 100 times faster than Hive. Spark uses Resilient Distributed dataset which are immutable, they are the data containers in which spark stores data [20].

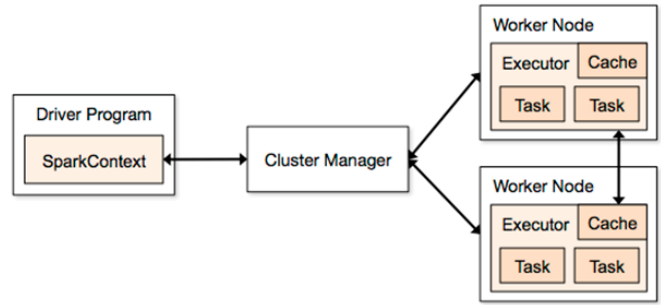

*Figure 4: Architecture of spark*

*h) Giraph*

Giraph is a tool specially designed for graphical analysis, it is an iterative graph processing system and provides scalability. It is an open source platform. For computation in giraph the input data must be a graph with vertices and directed edges. Giraph is being used by Facebook for analyzing social graphs [21].

*i) Apache Storm*

In traditional environment, we used to perform batch processing, but nowadays there is a great need for real time processing, and this functionality is provided by Apache Storm. It is open source, free of cost, fault tolerant, fast, scalable and reliable. Other functionalities provided by Storm are stream processing, online machine learning, ETL etc. [22]. It can easily connect with other querying and database technologies.

*j) Apache Flink*

Apache Flink is very much like Apache spark, both provides ML libraries, can run on top of Hadoop (HDFS, YARN) as well as can go in standalone mode, both provides stream processing as well as graph processing. The main difference between spark and flink is, in spark, Spark streaming wraps data streams into mini-batches, but in Flink streaming processes data streams as true streams, i.e., data elements are immediately "pipelined" though a streaming program as soon as they arrive [23]. This allows to perform flexible window operations on streams. It is even capable of handling late data in streams using watermarks.

*k) Zookeeper*

A centralized management system is required that can handle tools running in Hadoop environment, Zookeeper is there for this task, which was developed by Yahoo. It helps in configuration, availability, synchronization and coordination between different tools.

V. HYBRID MODEL FOR FACULTY RETRIEVAL

So many researches are going on in different fields in India but very few are known to everyone as there is no centralized system through which one can easy search for different researches and the faculties working on it, hence a centralized system is required. There is a need of efficient hybrid model for analysis. There is no such dataset available which contains information about different researches, the faculties working on it, universities in which these researches are going on and their location. We prepared our own dataset that contains information about all the NIT's in India. This dataset consists of attributes like: University Name, Faculty name, Faculties Designation, Research areas, Qualification, email ID to contact them, Department in which they work, Latitude and Longitude for the location purpose. With the help of this prepared dataset, we have tried to build an educational information infrastructure network that will help users to easily find out about the research works going under different faculties of NIT's, query about different research fields and to contact different people under same research area, one can also see how many faculties are there in different NIT's, which department has what number of faculties etc.

In our hybrid model, we used Spark as execution engine to provide faster and better performance, then we combined Hive with this execution engine as Hive can easily provide structure to our data and will help in querying and managing our data [24]. We used HDFS as distributed storage in our hybrid model, so we placed our data in HDFS first. To analyse our data we used one of the sparks library that is MLlib, we performed cluster analysis, in cluster analysis, Data within a cluster should be like other members of the cluster [20]. Our clusters were formed based on location i.e. latitude and longitude of the different universities. We choose k=3 i.e. number of clusters, so 3 clusters were formed and the nearby universities belonged to the same cluster.

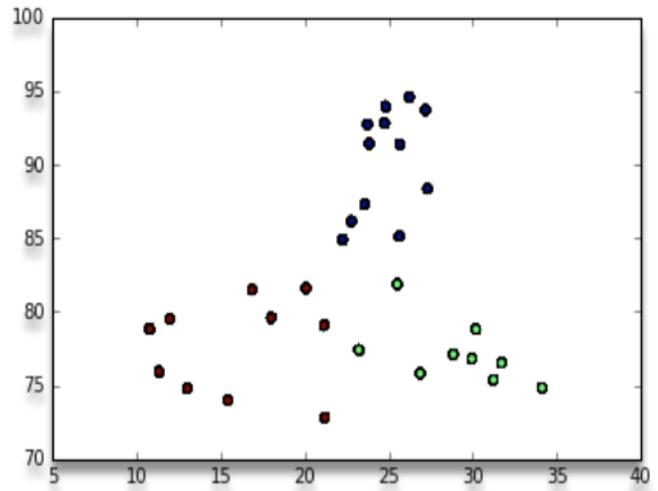

*Figure 5: Cluster Analysis on Spark*

As our data is already residing in distributed storage (HDFS), next we tried to manage and query our data, for this purpose we used Hive, to run things interactively in Hive we used beeline program [18]. Next, we created a blank table in Hive, named it 'faculty1' and define all attributes in the table. Then we load our data from HDFS to our table, by providing source path and destination name i.e. our table 'faculty1'. Now we are ready to perform queries on our data which is in our table in Hive. We performed queries like ordering by university name.

```
+---------------+-----------------------------+
|   university  |        faculty_name         |
+---------------+-----------------------------+
| NIT Warangal  | Dr. K KIRAN KUMAR           |
| NIT Warangal  | Dr. NAGA SRINIVASULU G      |
| NIT Warangal  | Dr. JOSEPH DAVIDSON M       |
| NIT Warangal  | SRI. SUBHASH CHANDRA BOSE P |
| NIT Warangal  | Dr. T. SADASIVA RAO         |
| NIT Warangal  | Dr. VASU V                  |
| NIT Warangal  | Dr. VENKAIAH N              |
| NIT Warangal  | Dr. Y. RAVI KUMAR           |
| NIT Warangal  | Dr. VEERESH BABU A          |
| NIT Warangal  | Dr. Adepu Kumar             |
| NIT Warangal  | Dr. SURESH BABU V           |
| NIT Warangal  | Dr. NARASIMHA RAO R         |
| NIT Warangal  | SRI. ASHOKKUMAR REDDY I     |
| NIT Warangal  | SRI. GUPTA G R K            |
| NIT Warangal  | SRI. VENKATESWARA RAO G     |
| NIT Warangal  | DR. G.AMBA PRASAD RAO       |
| NIT Warangal  | Dr. RAVI KUMAR PULI         |
| NIT Warangal  | DR. SRINADH K V S           |
| NIT Warangal  | DR. N . SELVARAJ            |
| NIT Warangal  | Dr. VENU GOPAL A            |
| NIT Warangal  | DR. GURURAJA RAO C          |
| NIT Warangal  | DR. NEELAKANTESWARA RAO A   |
| NIT Warangal  | DR. BANGARUBABU POPURI      |
| NIT Warangal  | DR. SRINIVASA RAO S         |
```

*Figure 6: Query showing results when ordered by university name*

Similarly, we can search in our database based on faculty names and/or their research fields. We have demonstrated some of the queries in this paper, the user can also perform other queries in diverse ways as per requirement.

```
0: jdbc:hive2://> SELECT university, faculty_name, research_area FROM faculty1 where research_area like concat('%','algorithm','%');
OK
17/04/23 00:05:11 [main]: WARN lazy.LazyStruct: Extra bytes detected at the end of the row! Ignoring similar problems.
+----------------+--------------------------+-----------------------------------------------------------------------+--+
|   university   |      faculty_name        |                              research_area                           |
+----------------+--------------------------+-----------------------------------------------------------------------+--+
| NIT Allahabad  | Dr. Deepak Kumar         | "Balanced Realization based frequency weighted model reduction algorithms |
| NIT Trichy     | Dr. P.Srinivasa Rao Nayak| Non- conventional optimization algorithms                             |
+----------------+--------------------------+-----------------------------------------------------------------------+--+
2 rows selected (0.228 seconds)
0: jdbc:hive2://> SELECT university, faculty_name, research_area FROM faculty1 where research_area like concat('%','data','%');
OK
17/04/23 00:10:30 [main]: WARN lazy.LazyStruct: Extra bytes detected at the end of the row! Ignoring similar problems.
+----------------+--------------------+---------------------------------------------------------------------+--+
|   university   |   faculty_name     |                              research_area                         |
+----------------+--------------------+---------------------------------------------------------------------+--+
| NIT Jamshedpur | Dr. Prakash Sarkar | Cosmology: Analysis of the Galaxy redshift survey data like SDSS |
| NIT Kurukshetra| Mahesh Pal         | "Classification and Feature selection with hyperspectral data    |
| NIT Delhi      | Dr. Jaya Thomas    | "Biodata Mining                                                   |
+----------------+--------------------+---------------------------------------------------------------------+--+
3 rows selected (0.137 seconds)
0: jdbc:hive2://> SELECT university, faculty_name, research_area FROM faculty1 where research_area like concat('%','architecture','%');
OK
17/04/23 00:11:10 [main]: WARN lazy.LazyStruct: Extra bytes detected at the end of the row! Ignoring similar problems.
+--------------+------------------------+-------------------------------------------------------------------------+--+
|  university  |     faculty_name       |                              research_area                             |
+--------------+------------------------+-------------------------------------------------------------------------+--+
| NIT Raipur   | Shaswat Sekhar Sarangi | "History of architecture                                                |
| NIT Warangal | DR. N.<ff>SUBRAHMANYAM | Distribution system studies; Standards for Distribution automation; Renewable energy integration studies; |
| NIT Hamirpur | Dr. Amitava Sarkar     | "Climate sensitive architecture                                         |
+--------------+------------------------+-------------------------------------------------------------------------+--+
```

*Figure 7: Results based on research fields*

```
0: jdbc:hive2://> SELECT university, faculty_name, research_area FROM faculty1 where faculty_name like concat('%','Prakash ','%');
OK
17/04/23 00:28:00 [main]: WARN lazy.LazyStruct: Extra bytes detected at the end of the row! Ignoring similar problems.
+------------------+-------------------------------+-----------------------------------------------------------------------+--+
|    university    |         faculty_name          |                              research_area                           |
+------------------+-------------------------------+-----------------------------------------------------------------------+--+
| NIT Patna        | Jyoti Prakash Singh           | "Sentiment Analysis
| NIT Patna        | Prakash Chandra               | "Heat Transfer
| NIT Raipur       | Mr. Satya Prakash Sahu        | "Artificial Intelligence & Expert System
| NIT Warangal     | Dr. Prakash Saudagar          | "Molecular and Biochemical parasitology
| NIT Warangal     | Dr.Prakash Saudagar           | NA
| NIT Bhopal       | Dr. Om Prakash Meena          | Communication Networks
| NIT Bhopal       | Dr. Jai Prakash Jaiswal<ff>   | "Development & convergence analysis of the iterative methods for solving nonl
| NIT Bhopal       | Dr. Jai Prakash Jaiswal<ff>   | NA
| NIT Jamshedpur   | Dr. Prakash Sarkar            | Cosmology: Analysis of the Galaxy redshift survey data like SDSS
| NIT Rourkela     | "Dr. Jaya Prakash Madda       | Associate Professor
| NIT Rourkela     | Dr. Parag Prakash Sutar       | "Drying and Dehydration
| NIT Rourkela     | "Prof. Dibya Prakash Jena     | Assistant Professor
| NIT Rourkela     | "Prof. Dibya Prakash Jena     | Assistant Professor
| NIT Rourkela     | Prof. Jyoti Prakash Kar       | "Thin Electronic Films
| NIT Rourkela     | Prof. Prakash Nath Vishwakarma| Low Temperature Condenser Matter Physics
| NIT Kurukshetra  | Sh.Prakash Chand              | NA
| NIT Kurukshetra  | Joy Prakash Misra             | "Machining Science
| NIT Jaipur       | Dr. Chetanya Prakash Sharma<ff>| "Physical Metallurgy
| NIT Agartala     | Dr. Suvra Prakash Mondal      | "Novel Sensors for Biomedical Applications
| NIT Silchar      | Dr. Jyoti Prakash Mishra      | Power Electronic Control in Electric Power and Energy Systems; Power Quality
| NIT Manipur      | Dr. Prakash Choudhary         | NA
| NIT Nagaland     | Dr. Prem Prakash Mishra       | "Operation Research
| NIT Uttarakhand  | Mr. Prakash Kushwaha<ff>      | NA
+------------------+-------------------------------+-----------------------------------------------------------------------+--+
23 rows selected (0.122 seconds)
```

*Figure 8: Results based on Faculty name*

Since visualization is an important aid to find out insights of your data, we used Tableau to visualize our data and its results [7]. Since we are dealing with big data, we connected tableau with Hive to take advantage of working in distributed environment [26, 27, 28].

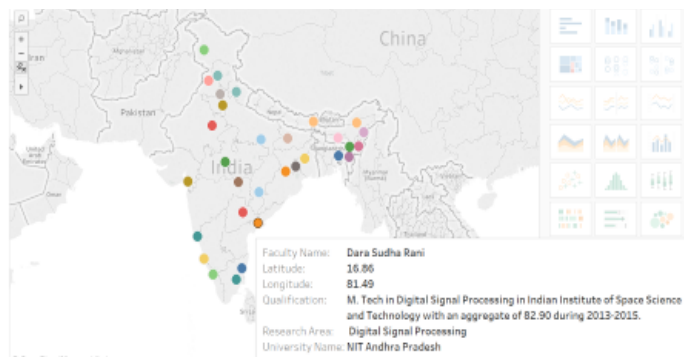

*Figure 9: Visualization of all NIT's India*

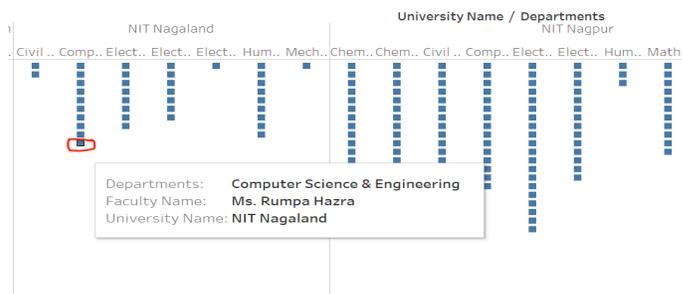

*Figure 10: Showing universities and the no. of faculties in each department*

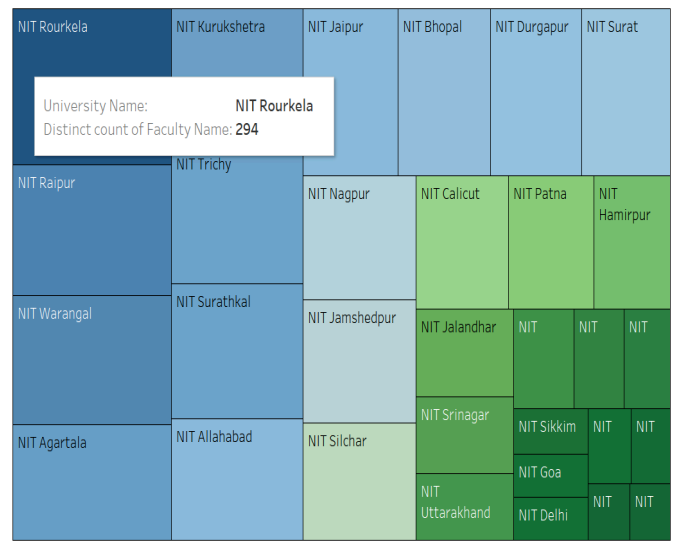

*Figure 11: Showing faculty number in each university*

## VI. CONCLUSION

This model provides easy and efficient way to find faculties and their research domains in India. As this model is built on distributed system, this tackles big data efficiently and can handle very huge dataset with ease. Tableau provides us the insight of the data as it is integrated with Hive, it too can process large dataset and give result in real-time.